\documentclass{article}
\usepackage{cite,url}
\usepackage{spconf}
\usepackage{algorithm,graphicx}
\usepackage{algorithmic}
\usepackage[usenames]{color}
\usepackage{amsfonts}
\usepackage{fancyhdr}
\usepackage{listings}%
\usepackage{arydshln}
\usepackage{amsmath,amssymb,amsthm}%
\usepackage{graphicx,psfrag,caption,subcaption}
\usepackage{listings}%
\usepackage{arydshln}
\usepackage{algorithmic}%

\DeclareMathOperator*{\diag}{\mathrm{diag}}

\def\xb{{\boldsymbol x}}

\def\Ab{{\boldsymbol w}}
\def\Xb{{\boldsymbol X}}
\def\Yb{{\boldsymbol Y}}

\def\hb{{\boldsymbol h}}

\def\vec{{\rm vec}}

\def\rank{\rm rank}

\def\Lb{{\boldsymbol L}}
\def\Cb{{\boldsymbol C}}

\def\ab{{\boldsymbol a}}
\def\bb{{\boldsymbol b}}

\def\qb{{\boldsymbol q}}

\def\yb{{\boldsymbol y}}

\def\Ab{{\boldsymbol A}}
\def\Bb{{\boldsymbol B}}

\def\Phib{{\boldsymbol \Phi}}
\def\Lambdab{{\boldsymbol \Lambda}}

\def\Psib{{\boldsymbol \Psi}}

\def\Ub{{\boldsymbol U}}

\def\ub{{\boldsymbol u}}

\def\Kc{{\cal K}}
\def\Vc{{\cal V}}
\def\Ec{{\cal E}}
\def\Gc{{\cal G}}

\def\Dd{\mathbb{D}}
\def\Rd{\mathbb{R}}
\title{Sampling and Reconstruction of Diffusive Fields on Graphs} 
\name{Siddartha Reddy and  Sundeep Prabhakar Chepuri \thanks{Software to reproduce the figures in this paper is available at https://ece.iisc.ac.in/$\sim$spchepuri/sw/graphdiffusionsampling.zip}}
\address{Indian Institute of Science, Bengaluru, India
}

\begin{document}
\ninept
\maketitle

\begin{abstract}
In this paper, the focus is on the reconstruction of a diffusive field and the localization of the underlying driving sources on arbitrary graphs by observing a significantly smaller subset of vertices of the graph uniformly in time. Specifically, we focus on the heat diffusion equation driven by an initial field and an external time-invariant input. When the underlying driving sources are modeled as an initial field or external input, the sources (hence the diffusive field) can be recovered from the subsampled observations without imposing any band-limiting or sparsity constraints. When the diffusion is induced by both the initial field and external input, then the field and sources can be recovered from the subsampled observations, however, by imposing band-limiting constraints on either the initial field or external input. For heat diffusion on graphs, we can compensate for the unobserved vertices with the temporal samples at the observed vertices. If the observations are noiseless, then the recovery is exact. Nonetheless, the developed least squares estimators perform reasonably well with noisy observations. We apply the developed theory for localizing and recovering hot spots on a rectangular metal plate with a cavity.
\end{abstract}
\vspace*{-1mm}
\maketitle

\section{Introduction}

Graph signal processing extends tools from classical signal processing to deal with data defined on networks and other irregular domains~\cite{ortega2018graph,shuman2013Emerging,sandryhaila2014big}.
We often come across such datasets in many diverse applications such as environmental sensing, traffic monitoring, mapping the human brain~\cite{huang2016graph}, cybersecurity~\cite{shah2010detecting}, and social networks, to list a few.

Similar to how we understand many physical phenomena by a partial differential equation that explains the evolution of a spatiotemporal field and relates it to the inducing sources, we can also understand the temporal evolution of data over a network or an irregular domain using a partial differential equation. For example, the heat equation is often used to model the traffic movement, infection or virus spread, or rumor propagation~\cite{chen2016detecting,shah2010detecting}. 

In this work, we focus on heat diffusion over networks. Specifically, we are interested in recovering diffusive signals on a graph by sampling a significantly smaller subset of vertices of the graph. This essentially amounts to localizing the underlying sources that drive the diffusion process from the observations that are collected at a few nodes. Oftentimes, the sources (e.g., traffic bottleneck or rumor sources) that induce the diffusion process are highly localized in the network or sparse in the vertex domain and hence are not usually bandlimited. Therefore, we require new sampling and recovery methods for graph signals that do not impose any structural or band-limiting constraints, unlike some of the existing graph sampling methods~\cite{chen2015discrete, chepuri2018graph,marques2015sampling}. Although band-limiting constraints are not needed for recovering the second-order statistics of a signal defined on a graph from the subsampled observations, the framework developed in~\cite{chepuri2017graph} cannot be used for localizing diffusive sources. Spatio-temporal sampling and reconstruction of diffusive fields on a regular domain under the assumption that the inducing sources are sparse are studied in~\cite{lu2009distributed,ranieri2011sampling}. Assuming that the underlying sources are known,~\cite{teke2017time} focuses on estimating the time instance when the sources appear. In contrast, we will assume that the start time of the sources are is known. 

In this work, we develop a graph sampling method to recover diffusive fields induced by an initial field and/or an external input that does not vary with time. The main results of this paper are as follows. When the underlying driving sources are modeled as an initial field or external input, we can localize and recover the sources by sampling a significantly smaller subset of vertices of the graph uniformly in time and by using a simple least squares estimator. To do so, we do not impose any constraints on the sources such as sparsity or bandlimitedness. Since we can compensate for the unobserved vertices with the temporal samples at the observed vertices, we can recover the sources without imposing any constraints. However, when the diffusion field is due to both the initial field and external input, to reconstruct the diffusive fields from the subsampled observations, we require either the initial field or external input to be bandlimited. If the observations are noiseless, then the recovery is exact. Nonetheless, the developed estimators perform reasonably well with noisy observations.

Throughout this paper, we will use upper (lower) case boldface letters to denote matrices (column vectors), and we will denote sets
using calligraphic letters.

\section{Graph signals}

Consider an undirected graph $\Gc = \{\Vc,\Ec\}$ with $N$ vertices (or nodes), where $\Vc = \{v_1,v_2,\ldots,v_N\}$ and  $\Ec$ represent the vertex set and edge set, respectively. Let us denote the graph Laplacian matrix associated with $\Gc$ as $\Lb \in \Rd^{N \times N}$. A graph signal is a function $x : \mathcal{V} \rightarrow \mathbb{C}$ with $x(v)$ being the value of the function at vertex $v \in \Vc$. Let us collect the function values $\{x(v_n)\}_{n=1}^N$ in a length-$N$ vector $\xb = [x_1,x_2,\ldots,x_N]^T$. 

For undirected graphs $\Lb$ is real symmetric, and hence admits an eigendecomposition $\Lb= \Ub \Lambdab\Ub^T$ with $\Ub = [\ub_1, \cdots, \ub_N]$ being the eigenvector matrix collecting the eigenvectors $\{{\boldsymbol u}_n\}_{n=1}^N$ and $\Lambdab = {\rm diag}[\lambda_1,\cdots,\lambda_N] $ being the diagonal matrix containing the corresponding eigenvalues $\{\lambda_n\}_{n=1}^N$. Here, $\diag[\cdot]$ refers
to a diagonal matrix with its argument on the main diagonal. The eigenvectors and eigenvalues of $\Lb$ provide the notion of frequency in the graph setting~\cite{shuman2013Emerging,sandryhaila2014big}. Specifically, $\{{\boldsymbol u}_n\}_{n=1}^N$ forms an orthonormal Fourier-like basis for graph signals with the graph frequencies denoted by $\{\lambda_n\}_{n=1}^N$.
The {\it graph Fourier transform} of $\xb$, denoted by $\xb_f$, is given by
\begin{equation}
\label{eq:GFT}
\xb_f = \Ub^T \xb \Leftrightarrow \xb = {\boldsymbol U} \xb_f.
\end{equation} 
We say that a graph signal $\xb$ is bandlimited, if its graph Fourier transform $\xb_f$ is sparse (i.e., contains a very few nonzero entries). Due to the uncertainty principle~\cite{bruckstein2002generalized}, a sparse graph signal $\xb$ is not bandlimited in general.

The frequency content of graph signals may be modified using {\it linear shift-invariant graph filters}~\cite{sandryhaila2013discrete} of the form
\begin{equation}
\label{eq:graph_fitler}
{\boldsymbol H}= {\boldsymbol U} \diag[{\hb_f}] {\boldsymbol U}^T  \in \Rd^{N \times N},
\end{equation}
where $\hb_f$ is the frequency response of the graph filter.

\section{Data model} \label{sec:prob}

Let us consider a signal $x(\Dd,t)$ in a physical domain $\Dd$ and temporal domain $t$.  We will assume that $x(\Dd,t)$ obeys the \emph{heat equation}
\begin{equation}
\frac{\partial x(\Dd,t)}{\partial t} = \alpha \nabla^{2}x(\Dd,t)+ q(\Dd),
\label{eq:heatequation}
\end{equation}
where $ \nabla^{2}$ is the Laplace operator, $\alpha$ is the diffusion constant, and $q(\Dd)$ is the external time-invariant input. When $t=0$, $x(0) = x(\Dd,0)$ represents the {\it initial field distribution}. Without loss of generality, from now on we will assume $\alpha = -1$. 

To solve such a differential equation on a surface or manifold, the manifold is discretized (e.g., using a Delaunay mesh), and the Laplace operator is replaced with a discrete Laplacian matrix (more specifically, a \emph{cotan-Laplacian} matrix) denoted by $\Lb$.  Thus, approximating \eqref{eq:heatequation} to 
\begin{equation}
\frac{\partial \xb(t)}{\partial t} = - \Lb \xb(t) + \qb,
\label{eq:heateqGraph}
\end{equation}
where $\xb(t) = [x_1(t),\ldots, x_N(t)]^T \in \mathbb{R}^N$ and $\qb = [q_1,\ldots, q_N]^T  \in \mathbb{R}^N$ are signals defined on the graph  represented by the Laplacian matrix $\Lb$. The differential equation \eqref{eq:heateqGraph} models heat diffusion on graphs, where the diffusion field is induced by $\xb(0)$ and $\qb$.
%
%In other words, $\xb(t)$ and $\qb$ are graph signals defined on the graph represented by the Laplacian matrix $\Lb$, and \eqref{eq:heateqGraph} models heat diffusion on graphs or networks. 

The solution to the non-homogenous differential equation \eqref{eq:heateqGraph} is given by~\cite{strang2015differentialeq} 
\begin{equation}
\begin{aligned}
\xb(t) &= e^{-t \Lb} \xb(0)+  \int_{0}^{t} e^{-s\Lb}\qb \,\,ds  \\
&= \Ub e^{-t \Lambdab} \Ub^T \xb(0)  +  \Ub \left(\int_{0}^{t} e^{-s\Lambdab} \,\,ds \right)\Ub^T \qb \\
&= \Ub e^{-t \Lambdab} \xb_{f}(0)  +  \Ub \left(\int_{0}^{t} e^{-s\Lambdab} \,\,ds \right)\qb_f
\label{eq:heateqSol}
\end{aligned}
\end{equation}
where $e^{\Lb} = \Ub e^{\Lambdab} \Ub^T \in \mathbb{R}^{N \times N}$ denotes the matrix exponential of $\Lb \in \mathbb{R}^{N \times N}$, $\xb(0)$ is the initial field distribution at $t=0$. Here, $\xb_{f}(0)  = \Ub^T\xb(0)$ and $\qb_f = \Ub^T\qb$ are, respectively the graph Fourier transforms of $\xb(0)$ and $\qb$. From \eqref{eq:graph_fitler}, we can see that the diffusive field $\xb(t)$ is obtained by filtering $\xb(0)$ and $\qb$ with graph filters having frequency responses $e^{-t \Lambdab}$ and $\int_{0}^{t} e^{-s\Lambdab} \,\,ds$, respectively.

Let us introduce the vectors  $\ab(t) =[e^{-\lambda_1 t}, \ldots,e^{-\lambda_N t}]^T$ and $\bb(t) = [f_t(\lambda_1), \ldots, f_t(\lambda_N)]^T$, where 
\[
f_t(\lambda) = \int_{0}^{t} e^{-\lambda s}\,\,ds = \frac{1-e^{-t\lambda}}{\lambda}
\]
with $f_t(0) = t$, and $f_0(\lambda) = 0$. We can now express \eqref{eq:heateqSol} compactly as
\begin{equation}
\xb(t) = \Ub \diag[\xb_{f}(0)] \ab(t)  +  \Ub \diag[\qb_f]  \bb(t). 
\end{equation}
Next, let us sample $\xb(t)$ uniformly in time at instances $\{t_k = \Delta  k, k=1,2,\cdots,T\}$ with step size $\Delta $  to obtain the data matrix $\Xb = [\xb(t_1), \xb(t_2),\cdots, \xb(t_T)] \in \mathbb{R}^{N \times T}$, which is given by
\begin{equation}
\Xb= \Ub \diag[\xb_{f}(0)] \Ab^T  +  \Ub \diag[\qb_f] \Bb^T,
\label{eq:heateqSol_vec}
\end{equation}
where $\Ab = [\ab(t_1), \ab(t_2),\ldots,\ab(t_T)]^T \in \mathbb{R}^{T \times N}$ and $\Bb = [\bb(t_1),\bb(t_2), \ldots,\bb(t_T)]^T \in \mathbb{R}^{T \times N}$. Also, let us observe a subset of $K$ out of $N$ mesh points and denote this subset with $\Kc \subseteq \Vc$, where $|\Kc| = K$. By introducing a selection matrix $\Phib \in \{0,1\}^{K \times N}$ that selects the field values at vertices indicated by $\Kc$, we can mathematically relate the subsampled observations to $\Xb$ as 
\[
\Yb = [\yb(t_1), \yb(t_2),\cdots, \yb(t_T)]  = \Phib \Xb.
\] 
In what follows, we will develop estimators to recover $\xb(0)$ and/or $\qb(0)$ from $\Yb$. 

\section{Diffusion field induced by $\xb(0)$ or $\qb$} \label{sec:init}
 In this section, we will develop a simple least squares estimator for reconstructing the diffusion field induced by $\xb(0)$ or $\qb$ from the subsampled data matrix $\Yb$. More importantly, we do not impose any band-limiting constraints on the sources. This means that the sources may be sparse in the vertex domain and can model localized events such as rumor or infection sources in a complex network, traffic accidents in a road network, or diffusion of hot spots on a surface, to list a few .  
  
 %\subsection{Diffusion field induced by $\xb(0)$ or  $\qb$}
Consider the case in which the diffusion field \eqref{eq:heateqSol} is induced by only the initial field $\xb(0)$ and the external input $\qb = {\bf 0}$. 
From \eqref{eq:heateqSol} and \eqref{eq:heateqSol_vec}, we have
\[
\Yb = \Phib\Xb= \Phib \Ub \diag[\xb_{f}(0)] \Ab^T.
\]
Vectorizing $\Yb$, we get a system of $KT$ equations in $N$ unknowns given by
\begin{equation}
\yb = \vec(\Yb) = \left(\Ab \circ  \Phib \Ub\right) \xb_{f}(0),
%= \left(\Ib \otimes \Phib\right) \left(\Ab \circ   \Ub\right) \xb_{f}(0)
\label{eq:obsx0}
\end{equation}
where $\circ$ denotes the Khatri-Rao (i.e., columnwise Kronecker) product, $\vec(\cdot)$ refers to the matrix vectorization
operator. Here, we have used the property $\vec(\Ab\diag[\bb]\Cb)= (\Cb^T \circ \Ab)\bb$. 

Suppose we choose $K$ and $T$ such that $KT \geq N$, and if the matrix $\Ab \circ  \Phib \Ub$ has full-column rank, then we can estimate $\xb_{f}(0)$ using least squares as
\[
\widehat{\xb}_{f}(0) = \left(\Ab \circ  \Phib \Ub\right)^\dag \yb,
\]
and localize the sources as
\[
\widehat{\xb}(0) = \Ub \widehat{\xb}_{f}(0).
\]
Using this in \eqref{eq:heateqSol} allows us to compute the diffusive field at any time $t$ and at all the vertices. When the diffusion field is induced by $\qb$ with $\xb(0)={\bf 0}$, the least squares estimator for $\qb$ may be developed along the similar lines.

%We end this subsection with the following remarks. 
The rank of the Khatri-Rao product of two matrices $\Ab$ and $\Bb$ (of appropriate dimensions) with no all-zero column satisfies~\cite{sidiropoulos2000uniqueness} 
\[
\rank (\Ab \circ \Bb) \geq \max\{\rank(\Ab),\rank(\Bb)\}.
\]
When the sampling time instances $\{t_1,\cdots,t_T\}$ and the eigenvalues of $\Lb$ are distinct, then the $T \times N$ Vandermonde matrix $\Ab$ will have full column rank of $N$ for $T \geq N$ and by construction does not have an all-zero column. Therefore, selecting rows of $\Ub$ such that there are no all-zero columns ensures that the rank of the matrix $\Ab \circ  \Phib \Ub$ will be $N$. In fact, observing only one node uniformly in time might result in the matrix $\Ab \circ  \Phib \Ub$ that has full column rank. However, in practice, depending on the observation time window, diffusion constant and the spectrum of  $\Lb$, $\Ab$ might be ill-conditioned. In such cases, $\Phib$ may be designed using sparse sensing (or sensor selection) techniques (e.g., see~\cite{chepuri2016sparse,ortiz2019sparse}) to obtain a full column rank matrix $\Ab \circ  \Phib \Ub$. 
\begin{figure}
	\centering
\includegraphics[width=0.5\columnwidth]{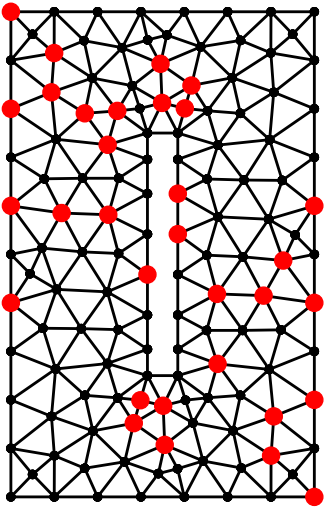}
\caption{Discretized metal plate with a cavity. The red (black) dots represent the observed (unobserved) vertices.}
\label{fig:mesh}
\end{figure} 

\begin{figure*}[!h]
        \centering
        	\psfrag{Celcius}{\scriptsize Celcius}
	        	\psfrag{true}{\scriptsize True}
			\psfrag{Estimated}{\scriptsize Estimated}
	\psfrag{Mesh #89}{\scriptsize $x(v_{89},t)$}
	\psfrag{Mesh #88}{\scriptsize $x(v_{88},t)$}
	\psfrag{Mesh #90}{\scriptsize $x(v_{90},t)$}
	\psfrag{Mesh #73}{\scriptsize $x(v_{73},t)$}
	\psfrag{Field diffusion [Celcius]}{\hskip-3mm\scriptsize Field diffusion [Celcius]}
	\psfrag{time}{\scriptsize time [s]}
	\psfrag{Source intensity}{\scriptsize Source intensity}
		\psfrag{Node index}{\scriptsize Node index}
	\psfrag{T=8}{\scriptsize $T=8$ }
	\psfrag{T=16}{\scriptsize $T=16$ }				
                      \begin{subfigure}[b]{0.3\textwidth}
                       	\centering	
                \includegraphics[width=0.8\columnwidth]{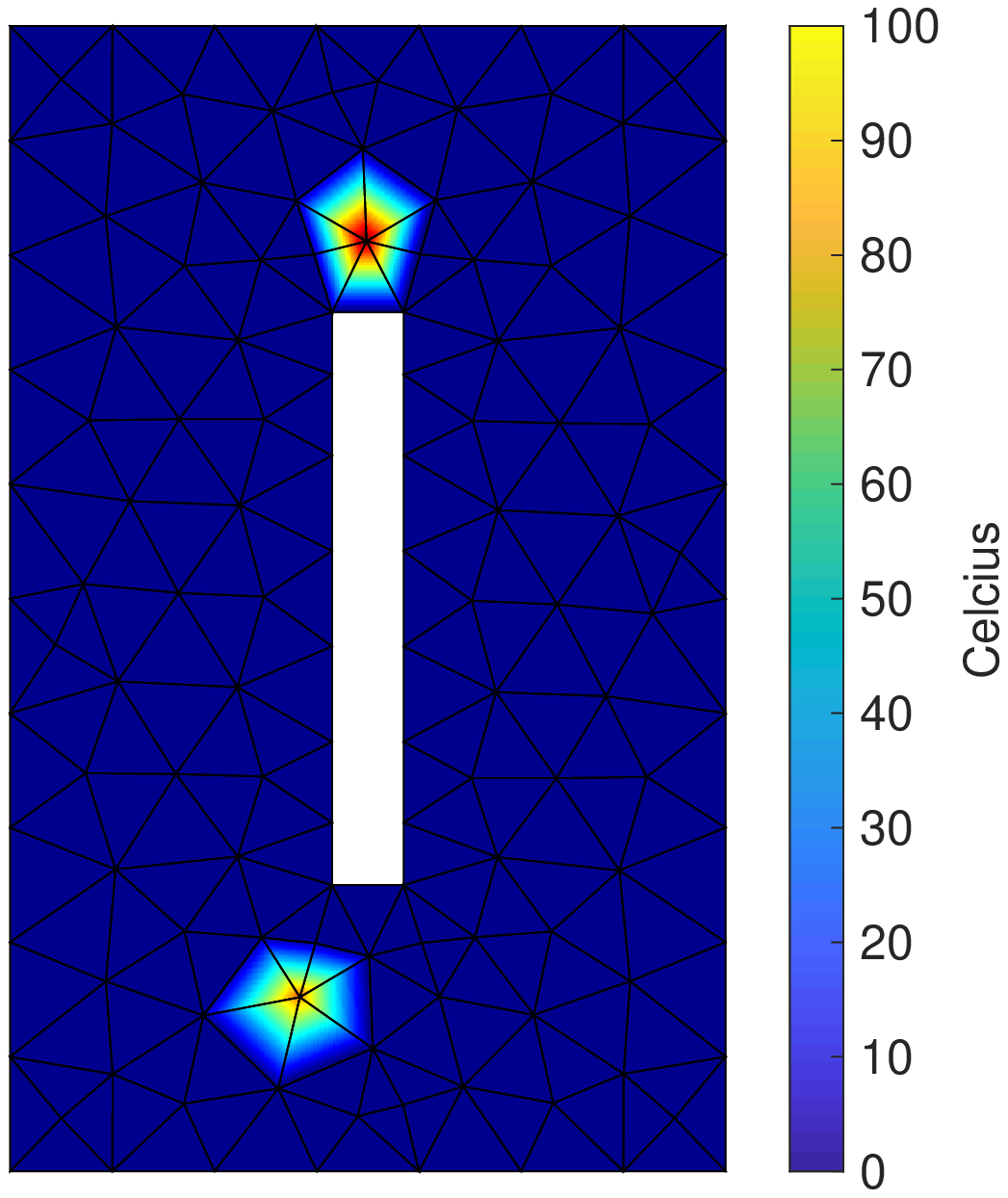}
                \caption{}
                \label{fig:x0}
        \end{subfigure}  
~  
 \begin{subfigure}[b]{0.3\textwidth}
	\centering
                \includegraphics[width=0.9\columnwidth]{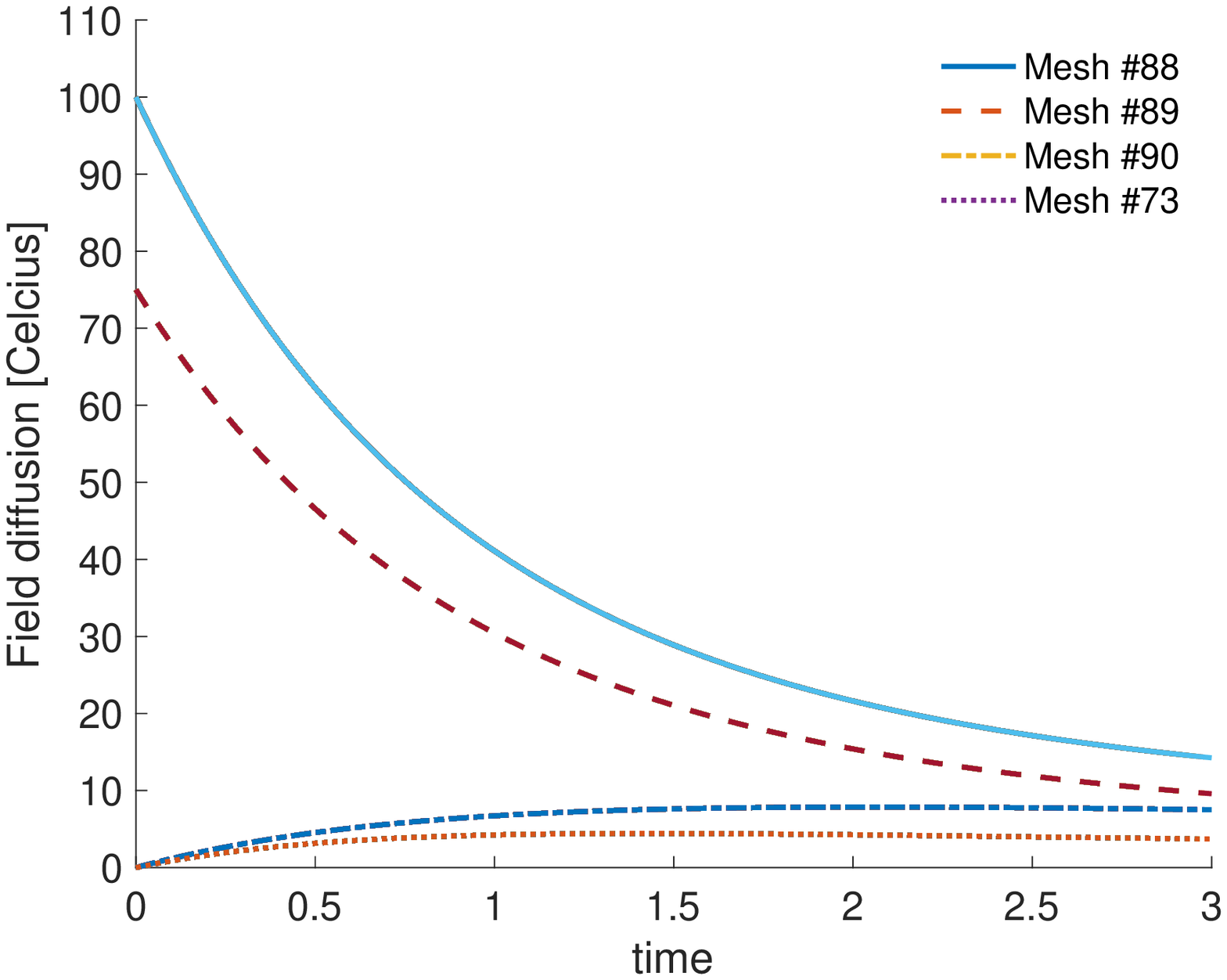}
                \caption{}
                \label{fig:xt1}
        \end{subfigure} 
       ~ 
 \begin{subfigure}[b]{0.3\textwidth}
	\centering
                \includegraphics[width=0.9\columnwidth]{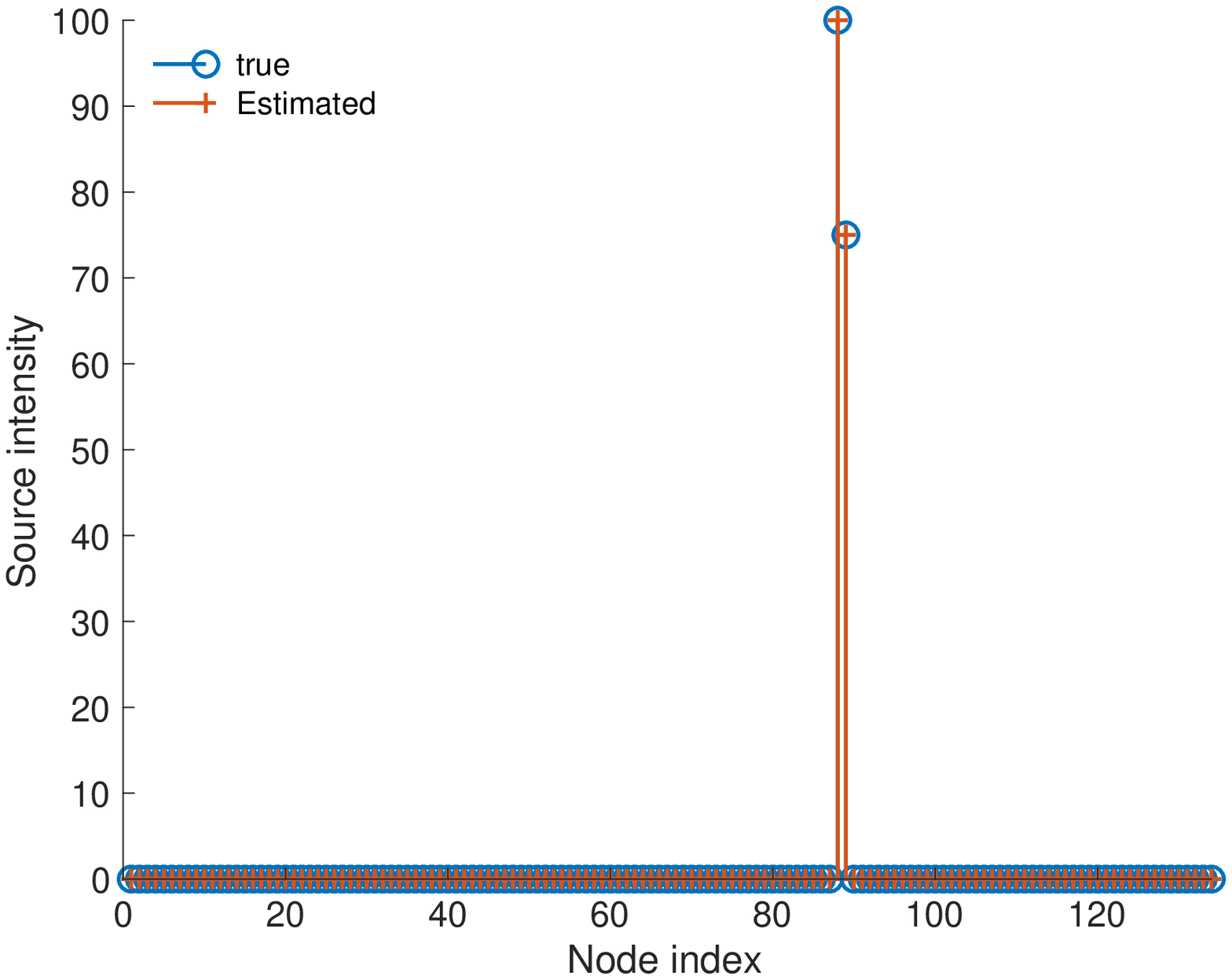}
                \caption{}
                \label{fig:x0hat}
        \end{subfigure}  
\\[1.5em]
 \begin{subfigure}[b]{0.3\textwidth}
	\centering
                \includegraphics[width=0.8\columnwidth]{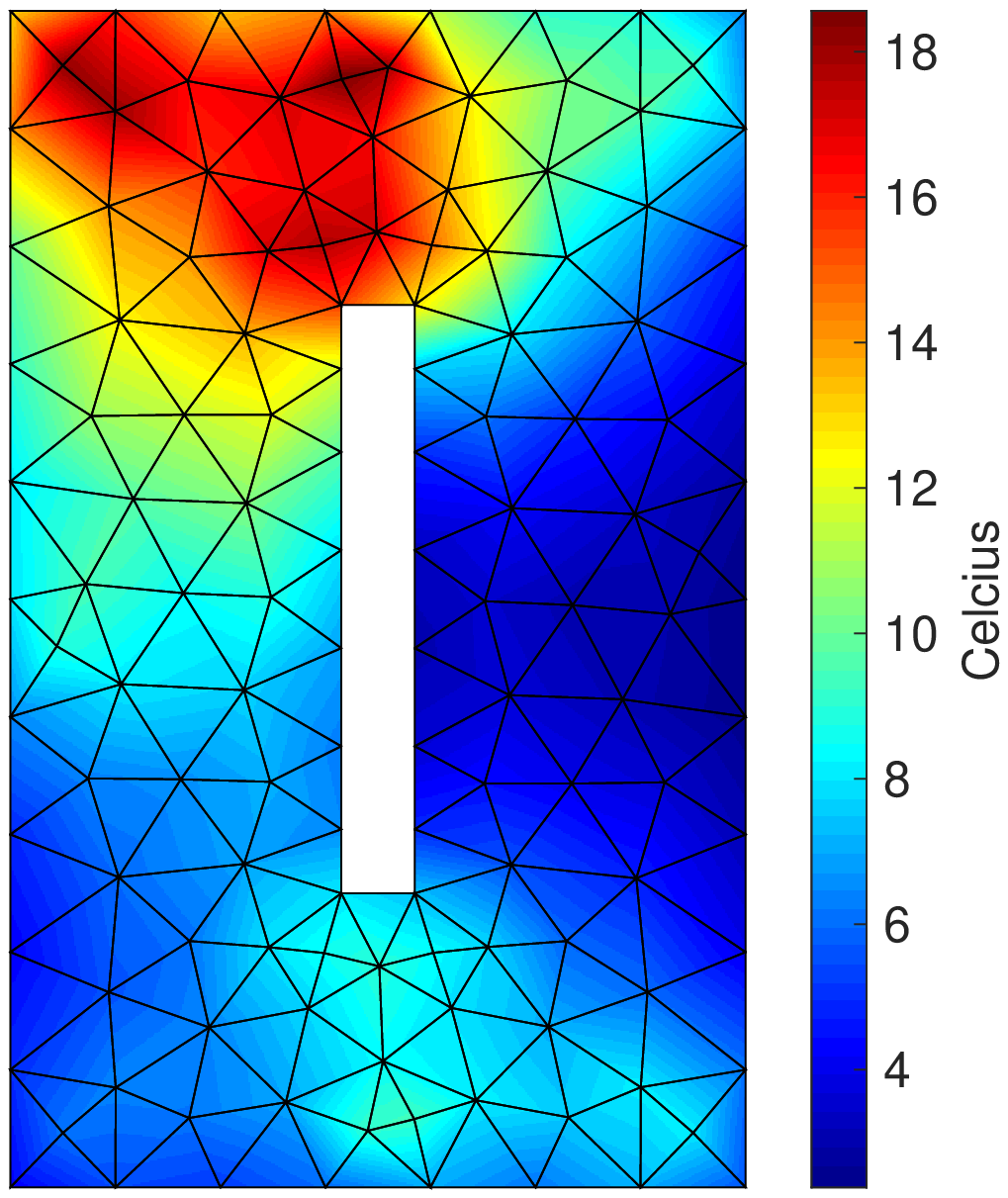}
                \caption{}
                \label{fig:q}
        \end{subfigure} 
~
 \begin{subfigure}[b]{0.3\textwidth}
	\centering
                \includegraphics[width=0.9\columnwidth]{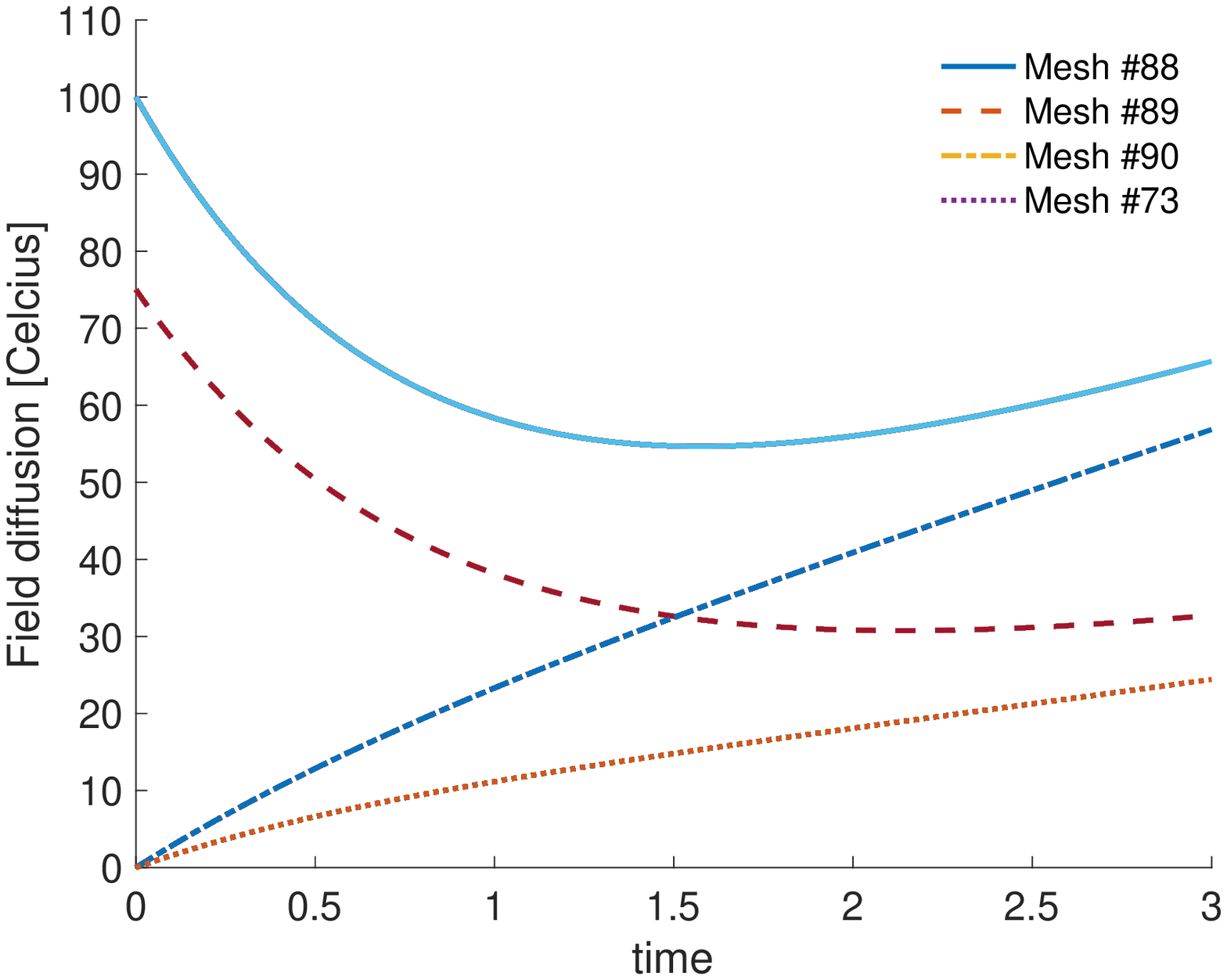}
                \caption{}
                \label{fig:xt2}
        \end{subfigure} 
~
 \begin{subfigure}[b]{0.3\textwidth}
	\centering
                \includegraphics[width=0.9\columnwidth]{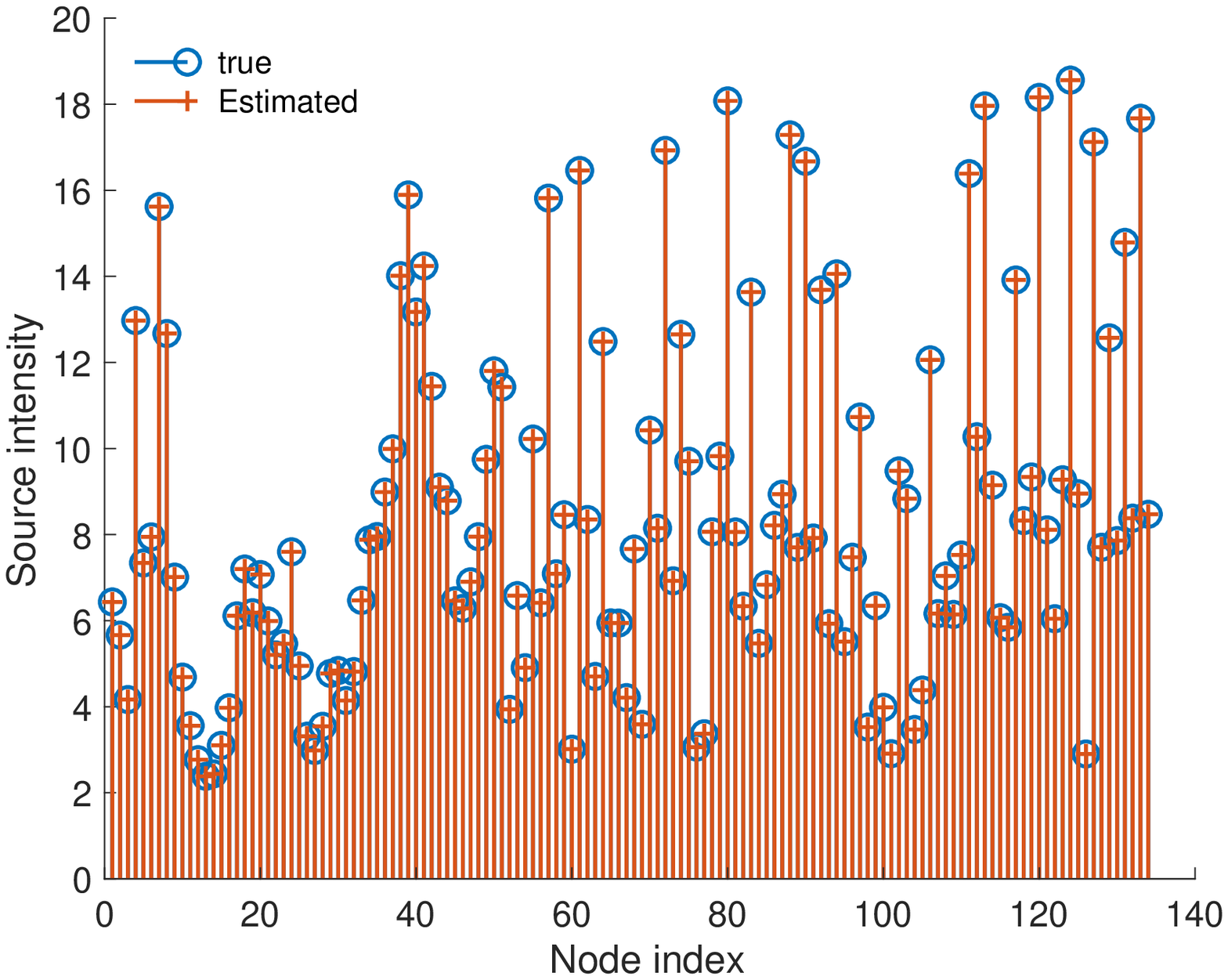}
                \caption{}
                \label{fig:qhat}
        \end{subfigure}       
        \caption{Recovery of diffusive fields. (a) Initial field distribution. (b) Evolution of the field with time when $\qb = {\bf 0}$. (c) Least squares based localization and reconstruction of $\xb(0)$. (d) Time-invariant and smooth external input $\qb$. (e) Evolution of field with time due to $\xb(0)$ and $\qb$. (f) Least squares based reconstruction of $\qb$.}                      
\end{figure*} 

\section{Diffusion field induced by $\xb(0)$ and $\qb$}
In this section, we consider the case in which the diffusion field is induced by $\xb(0)$ and a bandlimited time-invariant input $\qb$, and provide a simple least squares estimator to recover the underlying sources from the subsampled data matrix $\Yb$. Although we restrict $\qb$ to be bandlimited, we do not impose any band-limiting or other structural constraints on $\xb(0)$. The heat diffusion equation may be used to understand the movement of traffic in cities. Although the usual traffic movement may be assumed to be a smooth signal on a road network, there could exist a localized traffic bottleneck (e.g., due to an accident), which is a sparse non-bandlimited graph signal. Such diffusive fields may be modeled using \eqref{eq:heateqGraph} with a sparse $\xb(0)$ representing the localized events and a bandlimited 
$\qb$ representing the usual activity.

Recall that if $\qb$ is bandlimited, then $\qb_f$ will be sparse.Without loss of generality, let us assume that the first $P$ entries of $\qb_f = [q_{f,1},q_{f,2},\ldots, q_{f,N}]^T$ are nonzero. Then, the bandlimited (or smooth) signal $\qb$ may be expressed as a linear combination of the first few eigenvectors as
\begin{equation}
\label{eq:bl_q}
\qb = \sum_{i=1}^P \ub_i q_{f,i} = \Ub_P \qb_{f,P}, 
\end{equation}
where  $\Ub_P \in \Rd^{N \times P}$.

Vectorizing $\Yb$ in \eqref{eq:heateqSol_vec}, we have
\begin{equation}
\yb =\vec(\Yb) =\left[\begin{array}{cc}\Ab \circ   \Phib\Ub & \Bb \circ  \Phib \Ub \end{array}\right] \left[\begin{array}{c}\xb_{f}(0) \\\qb_{f}\end{array}\right].
\label{eq:vec_y_q}
\end{equation}
Substituting \eqref{eq:bl_q}, we get a linear system of $KT$ equations in $N + P$ unknowns
\begin{equation}
\begin{aligned}
\yb & = \left(\Ab \circ  \Phib \Ub\right) \xb_{f}(0)  + \left(\Bb \circ  \Phib \Ub\right) \Ub^T \Ub_P \qb_{f,P}\\
&= \left[\begin{array}{cc}\Ab \circ   \Phib\Ub & (\Bb \circ  \Phib \Ub)\Ub^T\Ub_P \end{array}\right] \left[\begin{array}{c}\xb_{f}(0) \\\qb_{f,P}\end{array}\right].
\end{aligned}
\label{eq:vec_y_blq}
\end{equation}
If the matrix $\Psib = \left[\Ab \circ   \Phib\Ub \quad (\Bb \circ  \Phib \Ub)\Ub^T\Ub_P \right]$ has full column rank, which requires $KT \geq N+P$, we can use least squares to obtain
\[
\left[\begin{array}{c}\widehat{\xb}_{f}(0) \\\widehat{\qb}_{f,P}\end{array}\right] = \Psib^\dag \yb,
\] 
and subsequently localize the underlying sources as
\[
\widehat{\xb}(0) = \Ub \widehat{\xb}_{f}(0); \,\,  \widehat{\qb} = \Ub_P \widehat{\qb}_{f,P}.
\]

In the previous subsection, we have seen that by appropriately selecting $K < N$ rows of ${\bf U}$ we may obtain a full column rank matrix $\Ab \circ  \Phib \Ub$ as $\Ab$ has full-column rank. As a consequence, by appropriately selecting $P > K$ rows of ${\bf U}$ will only increase the rank of $\left[\Ab \circ   \Phib\Ub \quad \Bb \circ  \Phib \Ub\right]$ by $N+P$. This means that we have to impose some structural constraint on $\qb$ to recover it uniquely from the subsampled data when the diffusive field is induced by both $\xb(0)$ and $\qb$. In other words, by sampling in time and observing all the $N$ nodes, we can recover $2N$ unknowns $\xb(0)$ and $\qb$ without any band-limiting constraints.

\section{Numerical experiments}

In this section, we apply the developed theory of graph sampling for reconstructing diffusive fields induced by hot spots on a metal block with a cavity. We use the \emph{partial differential equation} toolbox from MATLAB to mesh the surface. The generated mesh with $N=134$ vertices is shown in Fig.~\ref{fig:mesh}. We observe $K=32$ vertices uniformly in time in the interval $[0, 1.44] $s with step size $\Delta =  0.16$s and $T = 10$. The sampled vertices are also indicated in  Fig.~\ref{fig:mesh}. We present results for the following two cases: (i) diffusive field induced by $\xb(0)$, and (ii) diffusive field induced by $\xb(0)$ and $\qb$. 

\begin{figure}
		\psfrag{mse}{\hskip-6mm \scriptsize Normalized RMSE}
	\psfrag{sensors}{\scriptsize $K$ }
	\centering
\includegraphics[width=0.9\columnwidth]{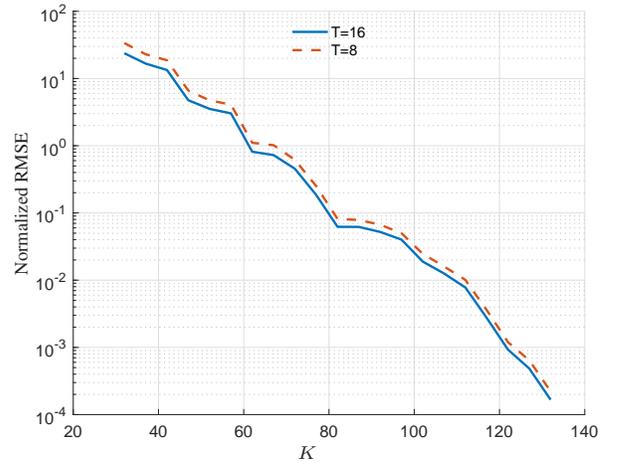}
\caption{Normalized root mean squared error for the diffusion field induced by $\xb(0)$ with $\qb = {\bf 0}$.}
\label{fig:rmse}
\end{figure}

As discussed in Section~\ref{sec:init}, to recover diffusive fields induced by $\xb(0)$ with $\qb = {\bf 0}$, we do not require any band-limiting constraints.  
To demonstrate this, for $\xb(0)$, we use a very sparse vector with only two non-zero entries at vertices $v_{88}$ and $v_{89}$. Since this initial field distribution is highly localized in the vertex domain, it is not bandlimited. Fig.~\ref{fig:x0} shows the initial field distribution at $t=0$, and Fig.~\ref{fig:xt1} shows the evolution of the diffusive field at vertices $v_{73}$, $v_{88}$, $v_{89}$,  and $v_{90}$ for different time instances. 
In Fig.~\ref{fig:x0hat}, we can see the exact localization of the hot spots in the noiseless setting using a simple linear least squares estimator, and more importantly, without using any sparsity constraints. In Fig.~\ref{fig:rmse}, we consider a noisy setting in which the observations in \eqref{eq:obsx0} are corrupted with Gaussian noise having zero mean and variance $10^{-5}$. We show the normalized root mean squared error (RMSE), averaged over 1000 independent Monte-Carlo experiments, for different values of $K$. Although the error decreases as $K$ increases, we can see that increasing $T$ beyond a certain value does not lead to better performance. This is because $\Ab$ becomes ill-conditioned as $T$ increases.

For the case in which the diffusion field is induced due to both $\xb(0)$ and $\qb$, we use a sparse $\xb(0)$ as before, and a bandlimited $\qb$ with $P=5$. Fig.~\ref{fig:q} shows the external time-invariant input, which is smooth on the surface. When $\qb \neq {\bf 0}$, we can see in Fig.~\ref{fig:xt2} that the field values do not decay with time as earlier. Fig.~\ref{fig:qhat} shows the exact recovery of $\qb$ using a simple linear least squares estimator (the reconstruction of $\xb(0)$ is similar to Fig.~\ref{fig:x0hat}, hence not shown), where we do not impose any sparsity constraints for recovering $\xb(0)$. As before, gathering more samples in time does not lead to better performance as both $\Ab$ and $\Bb$ become ill-conditioned, and as a consequence we need to sample more vertices.

\section{Concluding remarks}

In this paper, we discussed the sampling and recovery of diffusive fields on graphs induced by possibly non-bandlimited sources. When the diffusion field is induced by an initial field or a time-invariant external input, we can localize and recover the sources by sampling a significantly smaller subset of nodes uniformly in time without imposing any band-limiting constraints and by using a simple least squares estimator.  For diffusive fields induced due to an initial field and external input, we can exactly recover the sources from noiseless subsampled data when we constrain the external input  to be bandlimited. When the observations are noiseless, the recovery is exact. In essence, for diffusion models on graphs, we can compensate for the unobserved vertices with the temporal samples at the observed vertices.  

%\pagebreak

\bibliographystyle{IEEEtran}
\bibliography{IEEEabrv,journ,conf}

\end{document}